\documentclass{article}
\usepackage{emulateapj}
\usepackage{float,epsfig,psfig}
\usepackage{pstricks}
\def\hea4{{\it HEAO~A4}}
\def\heaoa2{{\it HEAO~A2}}
\def\heao1{{\it HEAO~1}}

\def\eg{{\it e.g.}~}

\def\h0{$H_{\rm o}=50$~km~s$^{-1}$~Mpc$^{-1}$}
\def\q0{$q_{\rm o}$}

\def\lsun     {$L_{\odot}$}

\def\etal    {{et~al.}~}

\def\cms3  {~{cm$^{-3}$}}

\newcommand{\mincir}{\raise
  -2.truept\hbox{\rlap{\hbox{$\sim$}}\raise5.truept \hbox{$<$}\ }}
\newcommand{\magcir}{\raise
  -2.truept\hbox{\rlap{\hbox{$\sim$}}\raise5.truept \hbox{$>$}\ }}

\begin{document}

\submitted{Submitted to ApJ January 3, 2003; in press v594 n1 ApJ September
  1, 2003} 
\title{Role of clusters of galaxies in the evolution of the metal budget in
  the Universe.} 
\author{A.~Finoguenov$^{1,2}$, A.~Burkert$^3$ and H.~B\"ohringer$^{1}$}
\affil{
{$^1$ Max-Planck-Institut f\"ur extraterrestrische Physik,
             Giessenbachstra\ss e, 85748 Garching, Germany}\\
{$^2$ Smithsonian Astrophysical Observatory, 60 Garden st., MS 3, Cambridge,
  MA 02138, USA}\\
{$^3$ Max-Planck-Institut f\"ur Astronomie, 
Koenigstuhl 17, 69117 Heidelberg, Germany}}
\authoremail{alexis@head-cfa.harvard.edu}

\begin{abstract}

Using the guidelines on SN element production provided by XMM-Newton, we
summarize the results of ASCA observations on the element abundance in
groups and clusters of galaxies. We show that while the metal production in
groups could be described by a stellar population with a standard local IMF,
clusters of galaxies require a more top-heavy IMF. We attribute an excess
heavy element production to an IMF evolution with redshift. Dating the
galaxy formation in clusters by observations of the star-formation rate, we
conclude that the IMF variations have occurred preferentially at $z\magcir
4$. We further combine our metallicity measurements with the mass function
of clusters to estimate the role of clusters in the evolution of the metal
content of the Universe. We argue that at no epoch stars are a major
container of metals, unless groups of galaxies are not representative for
the star-formation. This lends further support for the reduced (0.6 solar)
mass-averaged oxygen abundance in the stellar population.

\end{abstract}

\keywords{galaxies: intra-galactic medium; clusters: cosmology; cosmic
  star-formation} 

\section{Introduction}

The amount of baryons found in clusters of galaxies by X-ray observations is
similar to the amount of baryons locked in stars (Fukugita, Hogan, Peebles
1998, hereafter FHP). The high element abundance found in the cluster gas
makes cluster metals a significant entry in the total metal budget of the
Universe. Moreover, the flat radial abundance profiles of alpha-chain
elements (O, Mg, Si) in clusters suggest that the enrichment by SN type II
(which are responsible for 90\% of metal mass production, as opposed to only
25\% SN II contribution to Fe) occurred before the cluster collapse
(Finoguenov et al. 2000, 2001a, 2002b). Comparison with the predictions of
cluster formation in a low density Universe (as inferred by recent
observations) suggests that this enrichment should occur at redshifts
greater than 3. In comparison with the expectations for the total metal
budget at that time (e.g. Pagel 2002), clusters contain a significant
fraction of the metals, produced at high redshift. Direct observations at
high redshifts have not yet revealed the objects that account for metal
production associated with the observed star-formation rate density (Pettini
et al. 1999).

In this {\it Paper} we will consider a detailed analysis of the cluster
contribution to the metal balance in the Universe by combining the REFLEX
(B\"ohringer et al. 2001) results on the cluster mass function with ASCA
(Tanaka, Inoue, Holt 1984) results on the cluster metal abundance and a
preheating approach to explain the properties of groups of
galaxies. Suggesting cluster metals as an explanation for the missing metal
problem at high z, places the requirement on the top-heavy IMF needed to
explain the cluster metals at the redshift where such deviations have
recently been suggested from observations (Hernandez \& Ferrara 2002). 
Knowledge of IMF provides a key ingredient for understanding the
light-to-star-formation conversion, as well as the association of a stronger
feedback to high-redshift star-formation.

We adopt $H_{\rm o}=70$ km s$^{-1}$ Mpc$^{-1}$ to provide an unbiased
comparison between measured and theoretical values. The dependence of the
luminosity distance ($D_{L}$) on the deceleration parameter ($q_{\rm o}=0.5$
in our study) has a negligible effect (less than 5\%) on the results, as
this is a fossil record study of nearby systems.

\section{Data}

We collect the measurements of the ICM heavy element abundance, gas content,
and optical light for a sample of 44 groups and clusters from Finoguenov et
al.(2000; 2001; 2002a). Our sample constitutes $\sim70$\% of HIFLUGS, a
complete sample of bright clusters of galaxies and thus could be taken as
representative. Given the limited number of systems, we divide the sample
into three categories, that approximately correspond to a general census for
groups, poor and rich clusters.

A comparison among the systems is done at radii of equal gravitational
matter overdensity, which have been shown to correspond to similar
morphological mixes of galactic types and heavy element abundance content of
the ICM (Dressler \etal 1997; Cen \& Ostriker 1999; Finoguenov et
al. 2001a). This scaling is also preferred from the point of view of cluster
hydrodynamics (Evrard, Metzler, Navarro 1996). In view of our goal to
understand the metal budget, it is crucial to provide measurements at as
large radii as possible. Given the coverage of the clusters in our sample,
we select the radii of equal overdensity corresponding to $0.4r_{100}$
radius, where $r_{100}$ is a virial radius in a flat $\Omega_{m,\; \rm
o}=0.3$ $\Lambda$CDM universe (Pierpaoli, Scott, White 2001). For this
comparison the $r_{100}$ is estimated by rescaling the $r_{500}-T$ relation
of Finoguenov, Reiprich, B\"ohringer (2001b), assuming a NFW mass density
profile (Navarro, Frenk \& White 1996).

In Fig.\ref{m2l-fig} we show the properties of the sample. The sample is
binned and averaged according to the X-ray temperature, taken as a measure
of the mass of the system ($kT \sim M_{total}^{1.6}$, e.g. Finoguenov et
al. 2001b). The three bins, shown in Fig.\ref{m2l-fig} correspond to the
virial masses in the 0.3--1, 1--4, 4--14 $10^{14}M_\odot$ intervals. We plot
the total gravitational mass-to-light ratio, gas mass-to-light ratio, and
the values for $\Upsilon_{Fe}$ and $\Upsilon_{Si}$, where $\Upsilon_{Fe}$
and $\Upsilon_{Si}$ are Fe and Si mass to $L_B$ ratio, normalized to the
solar values of 0.00128 and 0.000702, respectively (calculated using the
meteoritic values from Anders \& Grevesse 1989).  We use the cumulative
values for gas, Fe, and Si mass, as well as optical light, obtained using
the spatially resolved measurements, as presented in Finoguenov et
al. (2000, 2001a,b). Deviations of the points around the mean value have
been studied by Finoguenov et al. (2001a). A significant spread of values
was found for the SN Ia products and was successfully interpreted as an
effect arising from the spread in the cluster formation epochs,
characteristic for the conventional low $\Omega_{m}$ cosmologies, assuming
the present-day SN Ia rate and the evolution of the SN Ia rate in agreement
with existing measurements (Finoguenov et al. 2001a and references
therein). At 0.4 $r_{100}$ the cluster gas reaches the 0.10 fraction of the
total mass, thus accounting already for more than 75\% of the primordial
baryon fraction with an additional $5-10$\% of the primordial baryon
fraction being presently locked in stars. Thus, we do not expect the gas
fraction to be by more than 20\% different at larger radii. Large amount of
observational data on outskirts of groups and clusters of galaxies
progressed our understanding on the non-gravitational effects (Tozzi \&
Norman 2001; Finoguenov et al. 2003a; Voit et al. 2003). Numerical
simulations that allow for a wide spread of SN II products at the epoch of
star-formation show that it does not prevent the gas from collapse on the
potential wells of systems studied here (e.g. Finoguenov et al. 2003a;
Tornatore et al. 2003). The lower fraction of gas reported here is
understood as a larger extent of the gas halo around the groups in the
virial units. Comparison of the distributions of the SN II and SN Ia
products in Finoguenov et al. (2000) in terms of element abundance and
mass-to-light ratio, suggests that the SN II products have been released
into the ICM before the cluster collapse, while SN Ia products trace well
the different spatial distribution between the gas and the stars inside the
cluster (see also De Grandi \& Molendi 2001). Since, as will be seen below,
early-type galaxies account for most of the star-formation that happened
inside the radius, where we observe the metals and the end of star-formation
in ellipticals precedes the formation of the cluster (Thomas, Maraston
Bender 2002), a release of the bulk of the SN II products should have
happened before the accretion shock heating in clusters changed the relative
gas-to-stellar mass distribution. To account for this difference (except in
Fig.\ref{m2l-fig}, where the observed values are plotted), we use the
presently observed gas mass in clusters to account for the contribution of
SN Ia, while for SN II, we take the 0.12\footnote{a somewhat larger value,
0.14, would be more consistent with the recent WMAP result (Spergel et al.
2003)} (precollapse) gas fraction to estimate the SN II contribution to the
Fe abundance (resulting in a factor of 1.8 higher SN II metal enrichment in
groups).

The release of the bulk of SN II products prior to cluster virialization is
a key assumption of the present work and all the conclusions derived here
rely on it. This implies a dominant role of early galactic winds in the
release of SN II products, as opposite to cluster environmental effects.
Modeling of the cosmic star-formation rate requires the ejection of energy
at early epochs ($z>4$) to prevent overproduction of stars (e.g. Kaiser
1986; Tornatore et al. 2003; Benson et al. 2003) and at $z\sim3$ to
reproduce the properties of the star-formation in early-type galaxies
(Tornatore et al. 2003; Oh \& Benson 2003).

%

\includegraphics[width=8.cm]{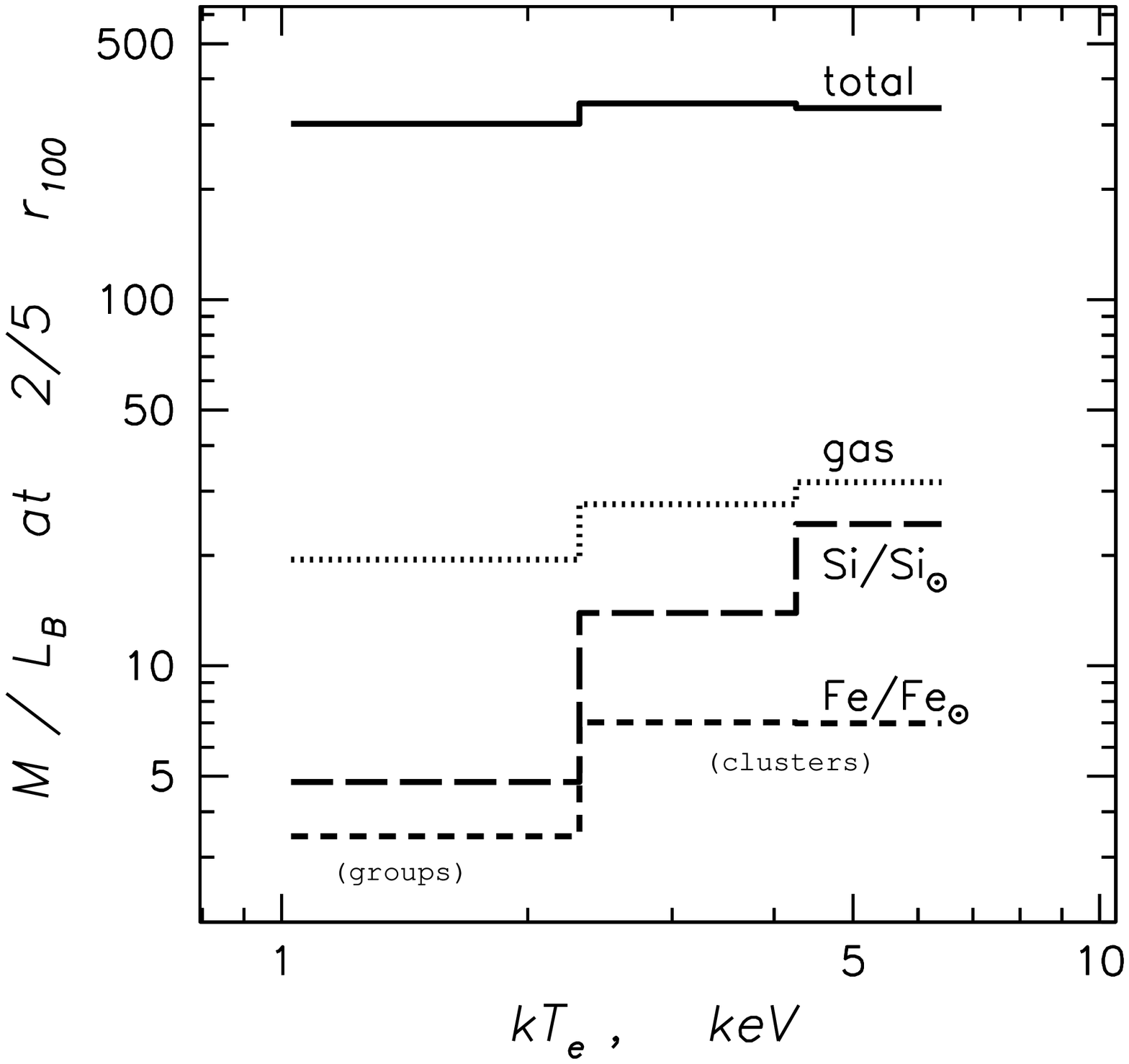}

\figcaption{$M$-to-$L_B$ ratios ($\Upsilon$) in groups and clusters of
galaxies vs the temperature of the ICM. The ratios of the cumulative values
within $0.4r_{100}$ are shown. The solid line indicates the ratios for the
total gravitational mass, dotted line -- for the gas. The short and long
dashed lines represent respectively the values for $\Upsilon_{Fe}$ and
$\Upsilon_{Si}$, expressed in solar units.
\label{m2l-fig}}

Galaxies can be subdivided onto disks and bulges, that have different
formation histories.  Since our key assumption on the early release of SN II
products is valid for bulges, but not valid for disks, we need to determine
the relative importance of bulges and disks. To obtain the morphological
fractions of galaxies, we relate the galaxy surface density $\Sigma$ in
units chosen by Dressler \etal (1997) to mass overdensity $\delta$ by
\begin{equation}\label{e:dr}
log_{10}(\Sigma)=0.58\times log_{10}(\delta)-0.40,
\end{equation}
using Eq.1 in Finoguenov et al (2000), the NFW profile, using the
measurements of radii of an overdensity of 500 from Finoguenov et al. 2001b,
and accounting for different assumptions in the Hubble constant. While we
use the Dressler relation established for local clusters, and it has been
shown by Helsdon \& Ponman (2003), that, accounting for the difference in
the projection to $\Sigma$ (which we implicitly do by adopting a scaling
with $\delta$), groups statistically obey the same relation. Using the
Eq.\ref{e:dr}, morphological fractions of E:S0:Sp within a radius of $0.4
r_{100}$ are found to be 0.27:0.46:0.27. To obtain the luminosity density
fractions, we multiply the morphological fractions by 1.9, 1.3, and 1.2 for
E, S0, and Sp galaxy types, correspondingly, and renormalize. These
conversion factors are taken from Arnaud \etal (1992) and were rescaled from
V to B band by multiplying the correction for Sp by 1.2 (\eg\ Postman \&
Geller 1984). The next step is to calculate the relation between the
observed light and the integrated initial mass of the stars, which we denote
as $M_{\rm *, o}$ (and its ratio to the present $L_B$ light as
$\Upsilon_{\rm *, o}$), assuming conversion factors of 6.5 and 1.5 for
bulges and disks, respectively (thereby assuming a Salpeter
IMF\footnote{such an assumption for $\Upsilon_{\rm *, o}$ correspond more to
a Scalo IMF for solar metallicity stars (Maraston 1998)}) as well as a disk
fraction in different galaxies as in FHP. This translates into the following
luminosity fractions of bulges and disks:
$L_B(spheroidal):L_B(disk)=0.74:0.26$ at 0.4 $r_{100}$, with the following
contribution to $\Upsilon_{\rm *, o}=4.8+0.4=5.2$.  We find therefore, that
the cumulative initial stellar fraction of total baryons amounts to 12\%,
similar to the K-band estimates (Lin, Mohr, Stanford 2003).
Recent observations of diffuse extra-galactic emission in Coma of
$L_B\propto 10^{11}$\lsun, suggests a contribution from IGSP (intergalactic
stellar population) of 20\% for the cluster total blue light (Gregg \& West
1998) and is therefore negligible. Part of this limit could be attributed to
the contribution of dwarfs.  In the model of Gibson \& Matteucci (1997), the
expected range of $\Upsilon_{\rm *,\, o}$ from dwarf galaxies lies in the
interval $2-6$, making a 10--20\% change with lower values more probable.

To estimate the amount of metals associated with the present-day stellar
population, we take $\Upsilon_{\rm *}=4$ for bulges, $\Upsilon_{\rm *}=1$
for disks, corresponding to the dynamical mass-to-light measurements
(e.g. van der Marel 1991), and a solar Fe abundance. We distinguish them
from the initial cumulative values by omitting the subscript 'o'.

\includegraphics[width=8.cm]{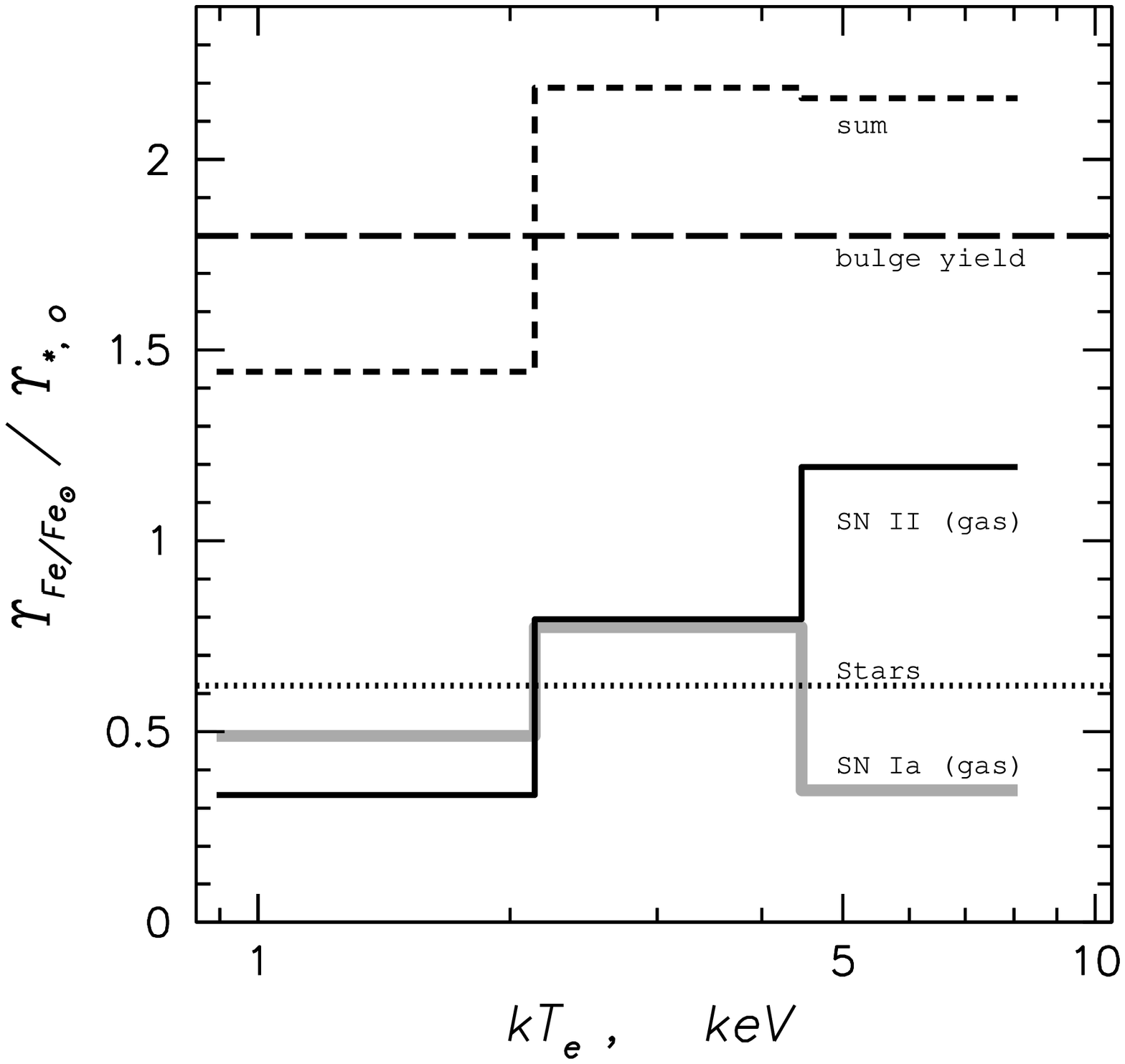}

\figcaption{The Fe enrichment schemes in units of stellar yields. The bulge
Fe enrichment amounts of 1.8 solar (Pagel 1987) and is shown by the
long-dashed line. The stellar component in clusters is shown by the dotted
line (assuming a solar value for the Fe abundance). The solid black (grey)
line indicates the SN II (SN Ia) contribution to the Fe enrichment in groups
and clusters. Cluster values are cumulative within a radius of 0.4 of
$r_{100}$.
\label{m2l-vir}}

%

%

To determine the relative enrichment from different supernova types, we
adopt the yields $y_{Si}=0.12M_{\sun}$, $y_{Fe}=0.05M_{\sun}$\ for a SN II
and $y_{Si}=0.15M_{\sun}$, $y_{Fe}=0.74M_{\sun}$\ for a SN Ia, confirmed by
analysis of clusters (Finoguenov et al. 2000). First XMM-Newton results on
the SN Ia enrichment in the cores of bright cluster galaxies revealed larger
Si/Fe ratio for SN Ia (Finoguenov et al. 2002b). However, with a larger
gas-to-light ratio this ratio was found to approach the W7 yields, assumed
here (Finoguenov et al. 2002b; Matsushita, Finoguenov, B\"ohringer
2003). Thus, for the region of study here, it is appropriate to assume the
W7 SN Ia yields. Changing the assumed values within the uncertainty of the
SN yields, does not change the separation onto SN II/Ia for clusters and has
a small effect for groups. The presented separation using the W7 model
maximizes SN II enrichment in groups. The uncertainty of the SN Ia and SN II
contribution to $\Upsilon_{\rm *,\, Fe}$ is 20\%. To guide the reader, the
highest Fe yield from SN II observed in clusters correspond to a SN II
contribution to the observed Fe abundance of 0.14 of photospheric solar
units of $4.68\times10^{-5}$ (0.20 in meteoritic solar units of
$3.23\times10^{-5}$) for iron number abundances relative to H.

In Fig.\ref{m2l-vir} we show the cumulative contributions from SN Ia and SN
II to Fe mass within 0.4 of $r_{100}$. As can be seen in Fig.\ref{m2l-vir},
that although the effective iron yield is nearly the same between groups and
clusters of galaxies and is similar to the bulge iron yield (Pagel 1987),
the SN II contribution increases by a factor of 3--4 between groups and
clusters of galaxies, which implies that there is no universal Silicon
yield. In the following we will conventionally express this result in terms
of the Oxygen yield, adopting the solar O/Si ratio from SN II in the ICM, as
observed by XMM-Newton (e.g. Finoguenov et al. 2002b).  The value for the
effective Oxygen yield ($\Upsilon_{\rm O}$/$\Upsilon_{\rm *,\, o}$) amounts
to 1.4 solar for groups and 4.1 solar for clusters, taking the measurements
of the SN II element abundance pattern in the cluster gas of Finoguenov et
al. (2002b). For comparison, the Salpeter IMF is characterized by one SN II
per 100 solar mass of stars. For O and Fe yields of $1.7M_\odot$ and
$0.07M_\odot$ per SN II, respectively, one finds a yield of 1.2 solar for O
and 0.4 solar for Fe matching our results for the groups.

In the calculation of the effective O yield we assume a stellar Oxygen
abundance of 0.6 solar, implied by the O abundance in early-type galaxies,
as seen in the XMM-Newton EPIC and RGS observations centered on bright
galaxies in groups and clusters (Peterson et al. 2002; Xu et al 2002;
Matsushita et al. 2003; Buote et al 2003), which is compatible with the
recent revision of the solar O abundance, which yields a value of
$0.56\pm0.06$ (Allende Prieto, Lambert, Asplund 2001). A similar value is
obtained from the chemical enrichment modeling of the LMC (Pagel \&
Tautvaisiene 1998). Higher values, however, are implied from the Mg
abundance measurements (Edmunds \& Phillipps 1997).  The apparent
contradiction between the measurements of the O yield from dwarf galaxies
and the O yield implied by the Milky Way chemical enrichment schemes has
been long known as an O yield problem (e.g. Pagel \& Tautvaisiene
1998). First XMM-Newton results also reveal a variation in the Mg/O ratio
(Matsushita, et al. 2003). All currently published XMM-Newton results on the
O/Mg ratio in ISM/ICM seem to fit into a scheme, where a subsolar O/Mg ratio
is characteristic of the stellar population (Xu et al. 2002; Matsushita et
al. 2003; Peterson et al. 2002; Buote et al. 2003) and a solar O/Mg ratio is
a characteristic of the ICM (Matsushita et al. 2003; Buote et al. 2003),
with a difference in the ratios being a factor of 2. In calculating the
metal budget, O plays a major role due to the high abundance of this element
and as we will discuss below, the conclusion whether most of the metals (by
mass) are in stars or gas, crucially depend on it. According to both X-ray
gas --- galaxy comparison and metal-poor vs metal rich stars, early chemical
enrichment has been characterized by an O/Mg ratio different by a factor of
two. Such a behavior of the O/Mg ratio, probably originating from different
averaging over the mass of SN II progenitors, makes it a tracer of the IMF,
independent of the total element production. However, in this picture the
Milky Way enrichment scheme is inconsistent with any scheme of early-type
galaxy formation and cluster enrichment. Currently there are two approaches
to model the end of star-formation in early-type galaxies, via ejection of
the gas (e.g. Matteucci 1994) or strangulation of the gas accretion (White
\& Rees 1978; Finoguenov et al. 2003a; Oh \& Benson 2002; Tornatore et
al. 2003). The first scenario predicts stellar metallicities to form before
or in parallel with the gas metallicities. The second scenario allows the
gas metallicity to form much ahead of the stellar metallicity, as preheating
of the gas surrounding the galaxy associated with its chemical enrichment
prevents this gas from accretion onto the galaxy causing a global feedback
effect. Thus, usage of the O/Mg ratio as a separate argument in favor of a
variable IMF supports the results only within a certain scheme of galaxy
formation.

\section{IMF evolution with the redshift}

As discussed in the previous section, there is a difference in the 
effective O yields implied by the analysis of groups and clusters of
galaxies.  So far all interpretations of the differences in the content of
SN II products between X-ray groups and clusters of galaxies focused on the
escape of metals from groups (Fukazawa et al. 1998; Finoguenov et
al. 2001a). High effective O yields, characteristic to clusters, cannot
be obtained with the standard IMF (e.g. Kroupa, Tout, Gilmore 1993),
appropriate for the star-formation processes observed at $z=0$. In the X-ray
regime we probe the systems covering two decades in mass
($10^{13}-10^{15}M_\odot$). With only one decade more, including the groups
with virial masses of $10^{12}M_\odot$ in the 2dF galaxy redshift survey,
Mart\'{\i}nez et al. (2002) account for most mass in the local
Universe. Thus, X-ray groups are a link between the clusters and the field
and a similarity in the chemical enrichment between groups and the field is
hardly by chance. Finoguenov, Briel, Henry (2003b) find that the chemical
enrichment of the X-ray filaments, which are a bona fide example of the
field environment, is similar to that of groups. Based on these arguments,
we consider variations in the IMF to occur only in massive clusters of
galaxies and associate such changes to the high-redshift epoch of
star-formation in clusters, following the theoretical work on the IMF
(Padoan, Nordlund, Jones 1997; Larson 1998).

In the hierarchical clustering model we expect a growth of galactic halos of
the same mass to occur earliest in the system of largest total mass, such as
a protocluster. This scheme has been strongly suggested by the tightness of
the scatter in the color-magnitude relation for cluster galaxies (Bower,
Kodama, Terlevich 1998). To estimate the redshift of the star-formation in
clusters we will use the extinction-corrected Madau plot in $\Lambda$CDM
Universe (e.g. Somerville, Primack, Faber 2001) and an assumption that the
star-formation density at high redshift is dominated by the formation of the
protocluster galaxies\footnote{Although, elliptical galaxies are thought to
form most of their stars via starburst, deviating from the original method
of Madau suitable for quiescent star formation, ellipticals should at least be
present in the Madau plot based on the IR band studies.}, followed by
proto-group galaxies and only then followed by the field. Numerical
simulations as well as considerations using the mean stellar ages imply that
the star-formation should decline at $z>5$ (e.g. Springel, Hernquist
2003). We adopt a decline of star-formation inversely proportional to z at
$z>3$, so at $z=6$ the star-formation rate is decreased by a factor of two
compared to $z=3$, which is consistent with current observational
restrictions, discussed in Springel, Hernquist (2003; see references
therein). We assume a linear rise in the star-formation rate with z at $z<1$
with a plato in the $1-3$ interval at the level of $0.07 M_\odot$ yr$^{-1}$
Mpc$^{-3}$. We note, that the conversion of luminosity density to the
star-formation rate for some of the methods rely on the assumption of the
Salpeter IMF, potentially leading to an overestimate of the star-formation
rate, which could be of some importance for most massive clusters, where IMF
variation are most strong.

\begin{equation}
\Omega_{*,\; \rm cluster}=\int_{t(z^{\rm cluster}_{\rm upper\; limit})}^{t(z=\infty)}\dot{\rho}_* (z(t)) dt / \rho_{crit}
\end{equation}

where we estimate $\Omega_{*,\; \rm cluster}$ using the results from a
complete X-ray survey of clusters of galaxies (REFLEX) on the cluster number
abundance (Reiprich \& B\"ohringer 2002), assuming stellar fraction of
baryons of 12\%, derived at $0.4 r_{100}$. Since the number abundance of
clusters is a strong function of their mass, as compared to changes in the
element abundance, we subdivided our sample onto four bins for further
analysis. In the following we will use the terminology group and cluster to
indicate trends with the mass of the system.

The above calculation provides an upper limit on the epoch of
star-formation, as we neglect the part of the star-formation proceeding in
parallel. To provide an estimate of the lower limit, we assume that most of
the star-formation in the systems we study happened at $z>2.5$,

\begin{equation}
\Omega_{*,\; \rm cluster}=\int_{t(z=2.5)}^{t(z^{\rm cluster}_{\rm lower\; limit})}
\dot{\rho}_* (z(t)) dt / \rho_{crit}
\end{equation}

At the peak star-formation rate of $0.07 M_\odot$ yr$^{-1}$ Mpc$^{-3}$ it
takes only one Gyr to reproduce the stellar population in the groups and
clusters of galaxies in the system mass range of the REFLEX sample, so the
choice of the lower limiting redshift is very important, e.g. at $z<2$
redshift-dependent effects could play no role in our sample. An independent
check of our choice of the star-formation redshift is given by the relative
importance of dwarf galaxies (Zabludoff, Mulchaey 2000), the tightness of
the color-magnitude relation (Bower et al 1998), consideration of the
feedback epoch (Finoguenov et al. 2002a), the delay between the formation
epochs of bulges and ellipticals, and studies of the star-formation in the
high-redshift progenitors of early-type galaxies (e.g. McCarthy et
al. 2001). These arguments predict the star-formation epoch for early-type
galaxies at around $z\sim3$, therefore supporting our estimate of the
redshift range. As was calculated above, recent episodes of star-formation,
as traced by disks, account only for 7\% of the total star-formation in
clusters.

\includegraphics[width=8.cm]{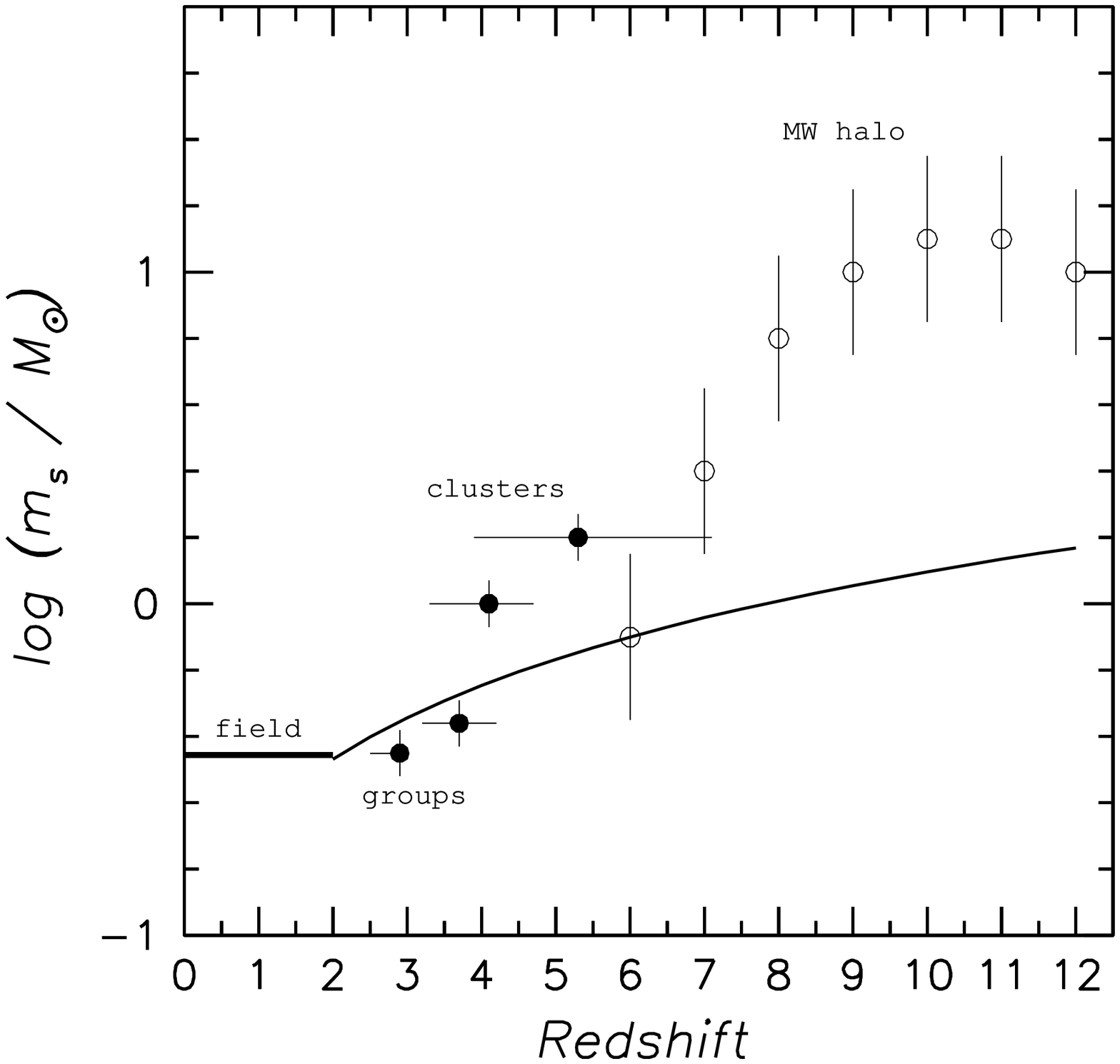}

\figcaption{Inferred values of the scale mass of the IMF ($m_s$), as a
function of redshift. Open circles denote the data from Hernandes \& Ferrara
(2001), filled circles denote the position of groups and clusters of
galaxies. The solid curve gives the redshift evolution of the Jeans mass of
cold star forming clouds, identified with $m_s$, resulting from the
temperature evolution of the cosmic microwave background.
\label{f:fer}}

We describe the IMF ($N(m)$) variation by the typical mass, $m_s$, given by
equation

\begin{equation}
 dN/dlog\, m \sim (1+m/m_s)^{-1.35}  
\end{equation}

from Hernandez \& Ferrara (2001). The value of $m_s$, which is identified
with the Jeans mass of cold star forming clouds, is judged from the amount
of metals produced by stars in the $12-40M_\odot$ range, compared to the
stellar mass in the $0.6-1.1M_\odot$ range. Such a choice of stellar mass
range is made to be consistent with definition of Hernandez \& Ferrara
(2001). For the low mass range, it is important to calculate the resulting
luminosity using the assumed IMF to weight the stellar tracks. The stars
below $0.6M_\odot$ do not contribute to the stellar light (e.g. Maraston
1998), so we do not consider them in our calculation.  We have checked that
using the $0.9-1.1M_{\odot}$ range does not strongly change our derived
values of $m_s$. Although, we do not calculate the light, the narrow range
of mass for the low mass stars allows us to neglect the effect of
differential averaging. As we have shown above, the amount of elements found
in groups corresponds closely to the Salpeter IMF, which in the formulation
of Hernandez \& Ferrara (2001) corresponds to $m_s=0.35M_\odot$. We preset
this mass scale for groups to provide an unbiased comparison to the data in
Hernandez \& Ferrara (2001).
The obtained values of $m_s$ for groups and clusters lie in
the $0.35-1.6M_\odot$ range, validating our simplified assumption on the
equal Fe yield per SN II.
We present the results of this analysis in Fig.\ref{f:fer}. Filled circles
indicate our data. The points derived from the modeling of the element
abundance in low-metallicity stars in the halo of the Galaxy by Hernandez \&
Ferrara (2001) are shown as open circles. One can see that the continuous
change of the IMF between groups and clusters of galaxies lies on the
continuation of the points derived from the low-metallicity stars. 
Faster-than-Salpeter fading of the light, deduced in LBG observations
(Ferguson, Dickinson, Papovich 2002) further supports our conclusion on the
biased IMF at high redshift ($z\sim6$). The assumption of a top-heavy IMF to
operate at high redshift is supported by the recent WMAP constraints on the
reionization epoch (Cen 2003).

Our results limit the occurrence of the biased IMF to $z>4$ and confirm the
findings of Hernandes \& Ferrara (2001) that changes in the IMF grow
exponentially with the redshift, exceeding the effect, predicted based on
the increase of the  CMB temperature (Padoan et al. 1997). The fact that the
enrichment of the intracluster gas is comparable to that of the halo stars,
argues against strong effects of the environment on the physics of the
star-formation.


Additional constraints on the assumed IMF shape, arise from the effects on
the dynamical mass-to-light ratios in galaxies (Zepf \& Silk 1996),
variations in the SN Ia rates with redshift (Smecker \& Wyse 1991), and C/O
and N/O ratios in the metal poor stars (Gibson \& Mould 1997). From the
point of view of X-ray observations of clusters, the introduction of a
varying IMF has only a 6\% effect on increasing $\Upsilon_*$ and 20\% on
$\Upsilon_{*,\;\rm o}$. Finoguenov et al. (2001a) showed that the observed
amount of SN Ia products in clusters could be understood if the release of
SN Ia products into the IGM occurs only when galaxies fall onto the cluster.
This scenario explains the lack of the SN Ia products in most massive
clusters. Alternative solution offered in Lin et al. (2003) consists in
indication from a more precise K-band light measurements that in the massive
clusters the stellar mass content is lower, that would solve the problem of
SN Ia products, while it would make a need for a top-heavy IMF even
stronger.

Consideration of C/O and N/O ratios should account for the variation in the
O yield in SN II chemical enrichment. With improvement in the sensitivity
and energy resolution at low energies, X-ray observations become sensitive
to the measurement of the C and N abundance and a proper comparison will
soon be possible.

\section{Metal content of the Universe}

To provide an insight into the role of clusters in the metal (all the
elements excluding H and He) budget of the Universe, we compare the metal
content in clusters with observations of other major metal entries. We use
the results from a complete X-ray survey of clusters of galaxies (REFLEX) on
the cluster number abundance (Reiprich \& B\"ohringer 2002) and combine them
with our metallicity measurements to calculate $\Omega_Z$. Detailed
measurements of the cluster mass and comparing the measurements using the
virial units, strongly reduces the amount of baryons associated with massive
virialized structures, compared to estimates of FHP. In addition, in
calculating the metal budget, we take into account a decrease in the
metallicity with decreasing mass of the system, substantially reducing the
implied metal budget, compared to estimates of Pagel (2002). We use an
estimate of the epoch of cluster enrichment, considered in the previous
section, which puts cluster metals in place at $z\sim3$ and implies no
evolution of the cluster metal content since then.  SN II are more frequent
in the Universe (SNe Ia:II is 1:7 in the Milky Way) and produce more metal
mass per event, compared to SN Ia. To go from Fe abundance to Z, we use our
separation into SN II and Ia contributions, take only the SN II part and
multiply it by the SN II metal-to-iron mass ratio. We take the SN II yields
from Nomoto \etal (1997), corrected for yields of S and Fe, as implied by
X-ray observations (e.g. Finoguenov et a. 2002b).  In this calculation O
makes up 75\% of metals, with Ne, Mg and Si contributing another 21\%. In
the calculation of the lower limit we ignore the mass of C and N, the
elements not yet measured in the ICM. We provide an upper limit on the metal
mass in clusters including C and N at solar ratios to O, which increases the
total mass of metals in clusters by 30\%.

\includegraphics[width=8.cm]{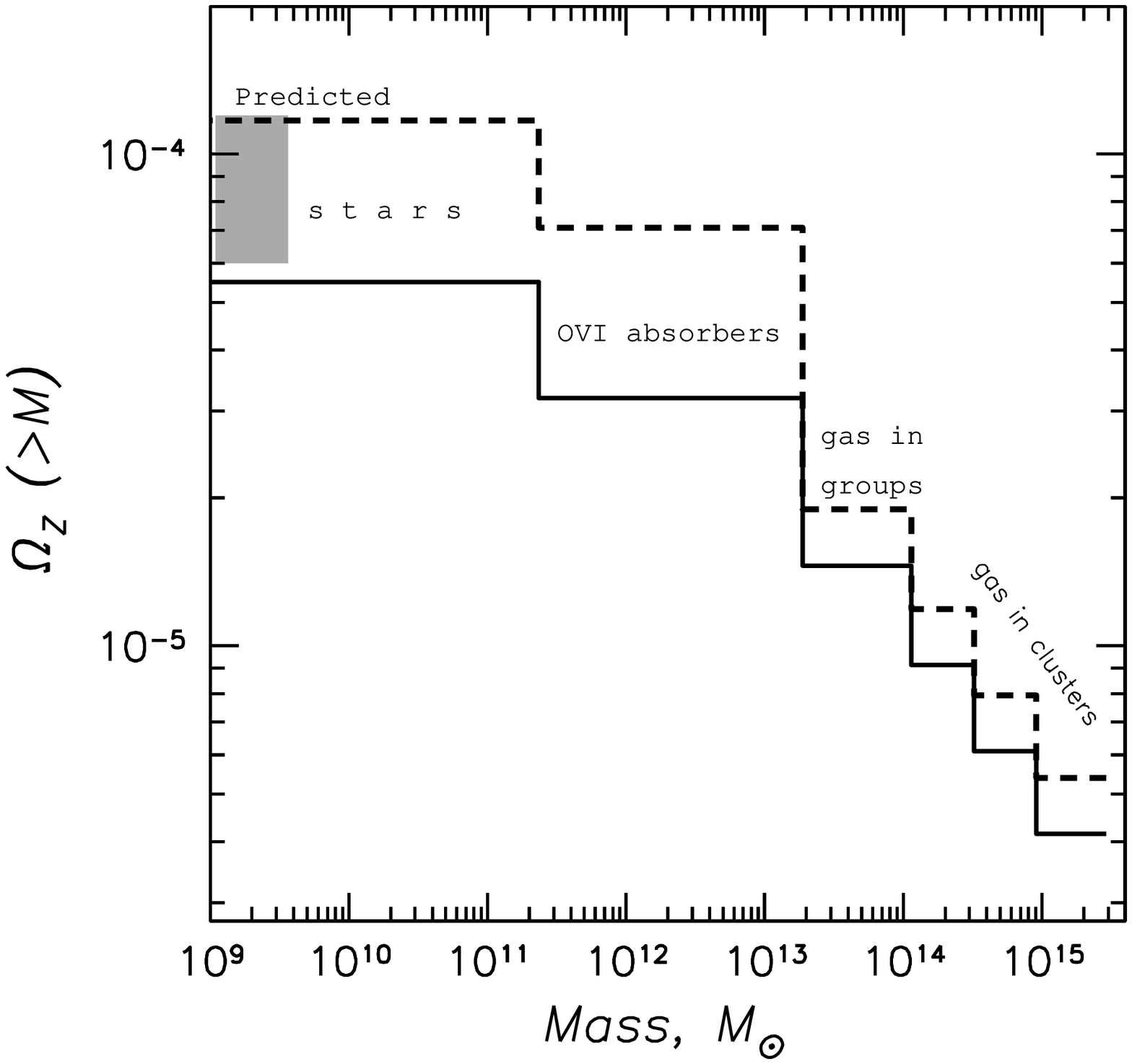}

\figcaption{Local cumulative $\Omega_Z(>M)-M$ relation. Solid and dashed
lines indicate a low and upper limit on the metal content, as explained in
the text and detailed in Tab.1. Grey shade is indicating the range of
predicted values.
\label{omegaz}}

Stars are known as an important metal entry at low redshift (Pagel 2002) and
as we have shown above, also at high redshift, yet at progressively lower
importance. To determine the $\Omega_Z$ of the stars at $z=0$, we take a
typical O abundance of 0.6 solar, as discussed above, and a 5--10\%
present-day stellar fraction of baryons in the field, obtained from the
K-band observations (Balogh et al. 2001; Huang et al. 2002). At $z=2.5$ we
take an estimate of Dunne, Eales, Edmunds (2002), scaling it by 0.6 for
self-consistency on the assumption of the O abundance in stars. We also
consider two major baryon reservoirs, $Ly_\alpha$ and OVI absorbers.
$Ly_\alpha$ absorbers account for most of the baryons at $z=2.5$,
contributing $0.3\times10^{-5}$ to $\Omega_Z$ (Pagel 2002). At $z=0$
$Ly_\alpha$ absorbers make up $32\pm6$\% of $\Omega_b$ (Penton et al. 2002;
McLin et al. 2002; Stoke et al. 2002). To estimate their metal contribution,
we assume that $Ly_\alpha$ absorbers at $z=0$ are a survived fraction of
their $z=3$ counterparts (Cen \& Ostriker 1999) and that their metallicity
does not evolve. To estimate the amount of metals associated with the OVI
absorbers, we use a lower limit by Tripp, Savage, Jenkins (2000) and a value
implied by ionization balance measurement of Mathur, Weinberg, Chen
(2002). No assumption on the O abundance is needed to calculate the
contribution of the OVI absorbers to the metal budget, while for the baryon
budget, we assume $0.14 O_\odot$ (Finoguenov et al. 2003b).  Both the metal
and baryon inventory are detailed in Tab.\ref{tab:m}.  

\vspace*{1cm}

{
\begin{center}
\footnotesize
\tabcaption{\centerline{\footnotesize
Inventory of metals and baryons in the Universe
\label{tab:m}}}
{\renewcommand{\arraystretch}{1.3}
\begin{tabular}{p{3.5cm}lccc}
\hline\hline
Component & $\Omega_{\rm Z}$, $10^{-5}$  & $\Omega_{\rm b}$, $10^{-2}$ & Refs\\
\hline                                                                    
\multicolumn{2}{c}{$z = 0$}\\
\hline                                                                    
Stars \dotfill & $2.3-4.6$ & $0.2-0.4$ & (1,2,3) \\
X-ray gas, clusters \dotfill & $1.4-1.8$ & 0.2 & (1) \\
OVI absorbers \dotfill & $1.8-5.2$ & $0.7-2.0$ & (1,8,9,10)\\
Ly$_\alpha$ forest \dotfill & 0.1 & $1.0-1.5$ & (6,7)\\
Total \dotfill & $5.6-11.7$ &  $2.1-4.1$ &  (1)\\
Predicted \dotfill & $6-12$ & 3.9 & (1, 5, 11)\\
\hline                                                                    
\multicolumn{2}{c}{$z = 2.5$}\\
\hline                     
Damped absorbers \dotfill & 0.1 & 0.1 & (5,11)\\
Ly$_\alpha$ forest \dotfill & $0.1-0.3$ & $1-5$ & (5,11)\\
protocluster gas  \dotfill & $1.4-1.8$ & 0.2 & (1)\\
ISM, dust \dotfill & $0.8-1.7$ & $0.1-0.2$ & (4,1)\\
Total \dotfill & $2.4-3.9$ & $1.4-5.5$ & (1)\\
Predicted \dotfill & $1.6-3.2$  & 3.9 & (1, 5, 11)\\
\hline               
\end{tabular}
}
\end{center}
References: \hspace*{0.3cm}{\footnotesize (1) this work; (2) Balogh et
al. 2001; (3) Huang et al. 2002; (4) Dunne, Eales, Edmunds 2002; (5) Pagel
2002; (6) Stoke et al. 2002; (7) Cen \& Ostriker 1999; (8) Tripp, Savage,
Jenkins 2000; (9) Mathur, Weinberg, Chen 2002; (10) Finoguenov et al. 2003b;
(11) Fukugita, Hogan, Peebles 1998}

}

\includegraphics[width=8.cm]{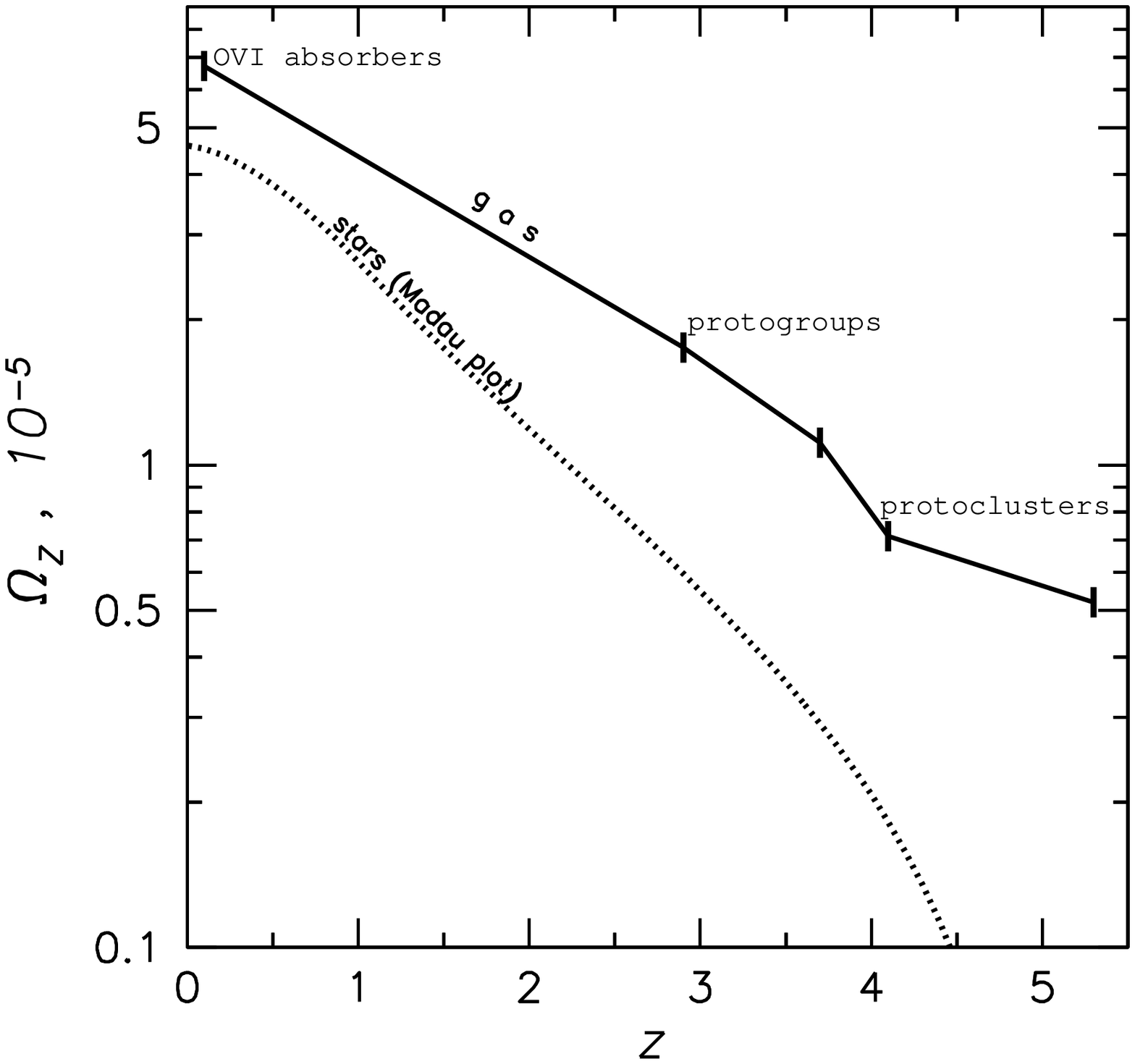}

\figcaption{Evolution of $\Omega_Z$ with redshift. Black line represents the
gas with labels indicating the major metal carrier at the redshift. The
dotted line shows the contribution to $\Omega_Z$ from the stellar
population. The Madau plot is normalized to the local K-band stellar baryon
content and a mass-average stellar oxygen abundance is assumed to be 0.6 solar.
\label{evl}}

The entries in Table \ref{tab:m} include the proposed solutions for both the
missing metals at high redshift (dust and/or warm gas in protoclusters) as
well as missing baryons in the local Universe (OVI absorbers; Ly$_\alpha$
absorbers). So within the uncertainty of measurements, neither baryons nor
metals are missing at either $z=0$ or $z=2.5$. In this comparison we do not
list the groups with masses below the detection threshold of the REFLEX
sample.  Part of these systems, that have lowest masses should be observed
as OVI absorbers and therefore we should not account for them twice. The
exact separation between X-ray emitting and OVI absorbing systems is not
entirely settled due to insufficient observational data on both sides. Since
we adopt an optimistic estimate of the amount of baryons, associated with
OVI absorbers, we consider it also as including the low-mass groups. There
could be some metals left unaccounted, if they are associated with the
groups, missed by both the OVI absorbers and REFLEX.

Fig.\ref{omegaz} shows the cumulative distribution of the metals over the
mass of the system and Fig.\ref{evl} outlines the evolution of various
components of $\Omega_Z$ with redshift. We specify two components, stellar
and gaseous. For stars we use the Madau-plot, normalizing it to the local
K-band estimate of the stellar content. For the metals in the gas we label
the dominant component at each epoch. At any epoch, metals are contained
mostly in the gas and not in the stars. This will no longer be true, once we
assume a solar O abundance for the stars and would indicate that groups of
galaxies are no longer representative of the star-formation processes.

\section{Conclusions}

We present the analysis of ASCA observations on the element abundance
measurements in groups and clusters of galaxies, determined at
$0.4r_{100}$. Using the results on the volume abundance of groups and
clusters of galaxies, provided by the REFLEX survey, we study the role of
metals associated with groups and clusters in the redshift evolution of the
metal content in the Universe.

\begin{itemize}

\item Detailed considerations of the amount of star-formation reveal
that the element production in groups is consistent with predictions for the
Salpeter IMF. Prevalence of the Salpeter IMF on the mass scales just a
factor of 10 smaller, argues in favor of the assumption made here that
groups do not loose SN II products.

\item The deviation in the chemical enrichment between groups and clusters
of galaxies, when described as an IMF variation with redshift, requires the
occurrence of strong IMF evolution at $z\magcir 4$.

\item The star-formation is guided by the IMF variation with redshift,
rather than environment, since our conclusions on the IMF evolution agrees
well with independent method of Hernandez \& Ferrara 2001, applied to the
Milky Way halo.

\item Proto-groups and proto-clusters of galaxies are an important reservoir
of metals at high redshift, and together with the dusty ISM of galaxies
provides a solution to the missing metal problem.

\item If star-formation in groups could be taken as representative, a low
(0.6 solar) mass-average O abundance in stars should be assumed. This
suggestion also finds support in many recent chemical enrichment studies
(e.g., Pagel \& Tautvaisiene 1998; Allende Prieto et al. 2001).

\end{itemize}

The importance of the above suggestions for a variable IMF at
high-redshift to early feedback and redshift-dependent corrections for the
star-formation rate, calls for direct confirmation of these results, that
will be available with XEUS, either in X-ray absorption or emission.

\section{Acknowledgments}

AF thanks Ralph Bender, Bernard Pagel, Cesare Chiosi, Andrea Ferrara,
Claudia Maraston, Francesca Matteucci, Trevor Ponman, Laura Portinari, Alvio
Renzini, Mike Shull, Daniel Thomas, and Simon White for useful discussions
and comments. The authors are thankful to Michael Loewenstein for the
comprehensive referee report. This work was supported in part by NASA Grant
16613323 and the Smithsonian Institution. AF acknowledges receiving
Max-Plank-Gesellschaft fellowship.


\begin{references}
\reference{} Allende Prieto, C., Lambert, D.L., Asplund, M. 2001 ApJ, 556, L63
\reference{} Anders, E., and Grevesse, N. 1989, Geochimica et Cosmochimica
Acta, 53, 197 
\reference{} Arnaud, M., Rothenflug, R., Boulade, O., Vigroux, L.,
 Vangioni-Flam, E. 1992, A\&A, 254, 49
\reference{} Balogh, M L., Pearce, F.R., Bower, R.G., Kay, S.T. 2001, MNRAS,
326, 1228
\reference{} B\"ohringer, H., Schuecker, P., Guzzo, L., Collins, C.A.,
Voges, W., et al. 2001, A\&A, 369, 826
\reference{} Bower, R.G., Kodama, T., Terlevich, A. 1998, MNRAS, 299, 1193
\reference{} Buote, D.A., Lewis, A.D., Brighenti, F., Mathews, W.G. 2003,
ApJ, subm. (astro-ph/0303054)
\reference{} Cen, R., Ostriker, J.P., 1999, ApJ, 519, L109
\reference{} Cen, R. 2003, ApJ, subm. (astro-ph/0303236)
\reference{} De Grandi, S., Molendi, S. 2001, ApJ, 551, 153
\reference{} Dressler, A., Oemler, A.,Jr., Couch, W.J., Smail, I., Ellis,
R.S., et al.  1997, ApJ, 490, 577
\reference{} Dunne, J.A., Eales, S., Edmunds, M.G.  2002, MNRAS, 335, 753 
\reference{} Edmunds, M.G., Phillipps, S. 1997, MNRAS, 292, 733
\reference{} Evrard, A.E., Metzler, C.A., Navarro, J.F., 1996, ApJ, 469, 494
\reference{} Ferguson, H.C., Dickinson, M., Papovich, C. 2002, ApJ, 569, L65
\reference{} Finoguenov, A., David, L.P., Ponman, T.J. 2000, ApJ, 544, 188
\reference{} Finoguenov, A., Arnaud, M., David, L.P. 2001a, ApJ, 555, 191
\reference{} Finoguenov, A., Reiprich T., B\"ohringer, H. 2001b, A\&A, 368,
749
\reference{} Finoguenov, A., Jones, C., B\"ohringer, H., Ponman, T. 2002a,
ApJ, 578, 74
\reference{} Finoguenov, A., Matsushita, K., B\"ohringer, H., Ikebe,  Y.,
Arnaud, M. 2002b, A\&A, 381, 21 
\reference{} Finoguenov, A., Borgani, S., Tornatore, L., B\"ohringer,
H. 2003a, A\&A, 398, L35
\reference{} Finoguenov, A., Briel, U., Henry, J.P. 2003b, A\&A,
submitted 
\reference{} Fukugita, M., Hogan, C.J., Peebles, P.J.E. 1998, ApJ, 503 518 
\reference{} Gibson, B.K., Matteucci, F. 1997, MNRAS, 291, L8
\reference{} Gibson, B.K., Mould, J.R. 1997, ApJ, 482, 98
\reference{} Gregg, M.D., West, M.J. 1998, Nature, 396, 549
\reference{} Helsdon, S.F., Ponman, T.J. 2003, MNRAS, 339, L29
\reference{} Hernandez, X., Ferrara, A. 2001, MNRAS, 324, 484
\reference{} Huang, J.-S., Glazebrook, K., Cowie, L.L., Tinney,
C. 2002, preprint astro-ph/0209440
\reference{} Kaiser, N. 1986, MNRAS, 222, 323
\reference{} Kroupa, P., Tout, C.A., Gilmore, G. 1993, MNRAS, 262, 545
\reference{} Larson, R.B. 1998, MNRAS, 301, 569
\reference{} Lin, Y.-T., Mohr, J.J., Stanford, S.A. 2003, ApJ, July 10 issue
(astro-ph/0304033)
\reference{} Maraston, C., 1998, MNRAS, 300, 872
\reference{} Mart\'{\i}nez, H.J., Zandivarez, A., Dom\'{\i}nguez, M.,
Merch\'an, M.E., Lambas, D. G. 2002, MNRAS, 333, L31
\reference{} Mathur, S., Weinberg, D., Chen, X. 2002, ApJ, submitted
(astro-ph/0206121)
\reference{} Matsushita, K., Finoguenov, A., B\"ohringer, H. 2003, A\&A,
401, 443
\reference{} Matteucci, F. 1994, A\&A, 288, 57
\reference{} McCarthy, P. J., et al. 2001, ApJ, 560, L131
\reference{} McLin, K.M., Stocke, J.T., Weymann, R.J., Penton, S.V.,
 Shull, J.M. 2002, ApJ, 574, L115
\reference{} Navarro, J.F., Frenk, C.S., White, S.D.M. 1996, ApJ, 462, 563
\reference{} Nomoto, K., et al. 1997, Nuclear Physics, A616, 79
\reference{} Oh, S.P., Benson, A.J., 2002, MNRAS, submitted, (astro-ph/0212309)
\reference{} Padoan, P., Nordlund, A.,  Jones, B.J.T. 1997, MNRAS, 288, 145
\reference{} Pagel, B.E.J. 2002, ASP CS, 253, 489
\reference{} Pagel, B.E.J. 1987, The galaxy, Dordrecht, D. Reidel Publ. Co.,
341  
\reference{} Pagel, B.E.J., Tautvaisiene, G. 1998, MNRAS, 299, 535
\reference{} Penton, S.V., Stocke, J.T., Shull, J.M. 2002, ApJ, 565, 720
\reference{} Peterson, J.R., Kahn, S.M., Paerels, F.B.S., et al. 2002, ApJ,
submitted, (astro-ph/0210662)
\reference{} Pettini, M., Ellison, S.L., Steidel, C.C., Bowen, D.V. 1999,
ApJ, 510, 576
\reference{} Pierpaoli, E., Scott, D., White, M. 2001, MNRAS, 325, 77
\reference{} Postman, M., Geller, M.J. 1984, ApJ, 281, 95
\reference{} Reiprich, T.H., B\"ohringer, H. 2002, ApJ, 567, 716
\reference{} Renzini, A., Ciotti, L., D'Ercole, A., Pellegrini, S. 1993,
ApJ, 419, 52
\reference{} Somerville, R.S., Primack, J.R., Faber, S.M. 2001, MNRAS, 320, 504
\reference{} Smecker, T.A., Wyse, R.F.G. 1991, ApJ, 372, 448
\reference{} Spergel, et al. 2003, ApJ, submitted, (astro-ph/0302209)
\reference{} Springel, V., Hernquist, L. 2003, MNRAS, 339, 312
\reference{} Stoke, J., et al. 2002, ApJ, submitted
\reference{} Tanaka, Y., Inoue, H., and Holt, S.S. 1984, PASJ, 46, L37
\reference{} Thomas, D., Maraston, C., Bender, R. 2002, Ap\&SS, 281, 371
\reference{} Tornatore, L., Borgani, S., Springel, V.,  et al. 2003, MNRAS,
in press (astro-ph/0302575)
\reference{} Tozzi, P., Norman, C. 2001, ApJ, 546, 63
\reference{} Tripp, T.M., Savage, B.D., Jenkins, E.B. 2000, ApJ, 534, L1
\reference{} van der Marel, R.P. 1991, MNRAS, 253, 710
\reference{} White S. D. M., Rees M. J. 1978, MNRAS, 183, 341
\reference{} Voit, G.M., Balogh, M.L., Bower, R.G., Lacey, C.G., Bryan,
G.L. 2003, ApJ, August 10 issue (astro-ph/0304447)
\reference{} Xu, H., Kahn, S.M., Peterson, J.R., Behar, E., Paerels,
F.B.S. 2002, ApJ, 579, 600
\reference{} Zabludoff, A.I., Mulchaey, J.S.  2000, ApJ, 539, 136
\reference{} Zepf, S.E., Silk, J. 1996, ApJ, 466, 114
\end{references}
\end{document}